\documentclass[manuscript]{aastex}
\usepackage{emulateapj5} 

\usepackage{amsmath}
\usepackage{amssymb}
\usepackage{url}

\newcommand{\Eq}[1]{Eq.~(\ref{#1})}

\shorttitle{Are perytons signatures of ball lightning?}
\shortauthors{Dodin and Fisch}

\begin{document}
\bibliographystyle{plainnat}

\title{Are perytons signatures of ball lightning?}

\author{I. Y. Dodin and N. J. Fisch}
\affil{Department of Astrophysical Sciences, Princeton University, Princeton, New Jersey 08543, USA}

\begin{abstract}
The enigmatic downchirped signals, called ``perytons'', that are detected by radio telescopes in the GHz frequency range may be produced by an atmospheric phenomenon known as ball lightning (BL). If BLs act as nonstationary radiofrequency cavities, their characteristic emission frequencies and evolution time scales are consistent with peryton observations, and so are general patterns in which BLs are known to occur. Based on this evidence, testable predictions are made that can confirm or rule out a causal connection between perytons and BLs. In either case, how perytons are searched for in observational data may warrant reconsideration, for existing procedures may be discarding events that has the same nature as known perytons.
\end{abstract}

\keywords{radiation mechanisms: general, waves}

\section{Introduction}

In the past several years, a number of unusual isolated signals were recorded with the 64-m Parkes Radio Telescope, Australia, in the frequency range $f \sim 1.2$-$1.5\,\text{GHz}$ \citep{ref:burke11, ref:bagchi12, ref:kocz12, phd:burke, arX:kulkarni14}. These signals, dubbed ``perytons'', exhibit a negative chirp $\dot{f} \sim -1\,\text{GHz}/\text{s}$ and last for hundreds of ms.\footnote{The duration of a whole signal must not be confused with the duration of its constituents in individual frequency channels of the telescope, which is typically tens of ms.} Also notably, perytons correlate with terrestrial settings such as time of day and weather and are detected in all, or most of, the 13 telescope beams. The common interpretation of the latter is that the signals are picked up by antenna sidelobes and thus must have large spectral flux densities, up to hundreds of kJy \citep{ref:burke11}. An alternative explanation could be (see below) that peryton sources are local and simply have a large enough angular size, namely, $\theta \gtrsim 1^\circ$. In either case, the signals are believed to have a terrestrial origin.

Identifying the specific sources of these signals yet remains an open problem. It is not entirely impossible that perytons are due to a man-made radiofrequency (RF) emission. However, this seems unlikely, because perytons cross the band 1.4-1.427 GHz, where terrestrial transmitters are legally forbidden to operate \citep{foot:esf}, and also exhibit amplitude modulation that, perhaps, excludes hardware failures as their origin \citep{ref:burke11, arX:khan14}. Thus perytons are more likely to be atmospheric phenomena. Yet, a specific mechanism through which the Earth's atmosphere produces such RF bursts remains elusive. (For most recent discussions, see \citep{arX:katz14, arX:khan14}.) Although perytons do correlate with weather, they are extremely rare compared to weather fluctuations and not necessarily accompanied by strong wind, rain, or thunderstorms \citep{ref:bagchi12, ref:burke2011b}. Perytons are therefore not likely to result from \textit{common} atmospheric phenomena. Rather, they may be emitted by structures, perhaps of decimeter size, that can last for about a second and change their geometry on the same time scale \citep{arX:katz14}. 

This paper will point out that, although exotic, such atmospheric structures are not unheard of; they are, in fact, widely known as the curious and equally puzzling phenomena called ball lightnings (BLs). We hence suggest that perytons are signatures of BLs. Although quantitative data on BLs is scarce, certain parallels between them and perytons are striking, with considerable circumstantial evidence linking these two types of effects. Based on this evidence, testable predictions are made that can confirm or rule out a causal connection between perytons and BLs. Note, however, that our argument does not rely significantly on any specific model of BL, which are numerous \citep{ref:smirnov93}. In this sense, our paper is not about BLs; rather it builds on some of already existing theories of BL to explain the origin and properties of perytons. We also identify a selection bias in peryton observations and suggest that it is hindering the search for physical mechanisms that could be responsible for generating perytons.

\section{Perytons vs ball lightnings}

\subsection{Frequency range} 

BLs are believed by many authors, although not unanimously \citep{ref:smirnov93}, to be accompanied by RF activity in just the frequency range where perytons are observed. Indeed, it has been suggested that a BL can serve as a natural electromagnetic cavity \citep{ref:kapitsa55, ref:watson60, ref:tonks60, ref:silberg61, ref:dawson69, ref:jennison73, ref:endean76, ref:muldrew90, ref:zheng90, ref:wesselberg03}. The lowest eigenmode of such a cavity has frequency\footnote{A geometric factor of order one \citep{ref:kapitsa55} is omitted because $D$ itself varies within almost two orders of magnitude \citep{ref:smirnov93}, and perytons too may exist also beyond the frequency range within which they are being presently studied.}
\begin{equation}\label{eq:fce}
f_c \sim c/D,
\end{equation}
where $D$ is the BL diameter, and $c$ is the speed of light. Even if the radiation were well-trapped inside the cavity, one can still expect it to somewhat radiate at frequency $f_c$. Typically, $D \sim 20\,\text{cm}$ \citep{ref:smirnov93}, so $f_c \sim 1.5\,\text{GHz}$, which is close to peryton frequencies.

Measurements of the RF emission generated naturally at thunderstorms also support the theory that BLs represent cavity phenomena. As shown in \citep{ref:kosarev68, ref:kosarev70}, the spectral density of this emission (measured at discrete frequencies) increases with frequency at $f \gtrsim 1.0$-$1.3\,\text{GHz}$, in striking contrast with the spectral density in the sub-GHz range, which decreases with $f$. It was suggested that this GHz radiation may be associated with BLs \citep{ref:kosarev68, ref:kosarev70}. Note also that the signals in individual frequency channels recorded in those studies are similar to the corresponding signals recorded for perytons.

\subsection{Frequency chirping} 
\label{sec:chirp}

The only reported quantitative observation of a natural BL \citep{ref:cen14} shows that the BL size can evolve significantly on a fraction of a second. During the quasistationary phase of the BL, this size, in fact, increased at the rate $\dot{D}/D \sim 0.5\,\text{s}^{-1}$. The value of $D$ itself cannot be inferred directly from the observations in \citep{ref:cen14}, which were performed from a large distance (0.9~km) and, as the authors pointed out, gave only the ``apparent'' diameter (in the several-meter range) rather than the actual diameter of the BL. But if one estimates $D$ to be 20\,cm, as usual, this leads to $\dot{f}_{\rm BL} \sim -0.75\,\text{GHz}/\text{s}$. Again, this value is consistent with what is seen for perytons. 

Let us now discuss whether the expanding-BL model explains the characteristic \textit{shape} of $f(t)$ observed for perytons. For this, a brief excursion into the history of peryton studies is needed. Perytons were originally discovered during an archival data survey \citep{ref:burke11} inspired by the discovery of the so-called Lorimer burst (LB), a similarly-shaped chirped GHz signal reported in \citep{ref:lorimer07}. As opposed to perytons, the LB was observed in only three beams of the Parkes antenna and thus was identified as a signature of a distant, extraterrestrial event associated with a few-ms RF emission. Such an RF signal undergoes dispersive spreading when propagating in space plasma. Specifically, its instantaneous frequency, as detected after time $t$ at a given distance $\ell$ from the source, satisfies~\citep{arX:katz14}
\begin{equation}\label{eq:ch}
 \frac{d}{dt}\, \big[f^{-2}(t)\big] = C(\ell),
\end{equation}
where the time-independent $C(\ell)$ is determined by the plasma density integrated along the signal trajectory (also known as the ``dispersion measure''). Choosing the value of $C(\ell)$ to fit the observations places the LB origin outside our galaxy. This motivated the search for other signals that would be similar to the LB, and, through that, perytons were discovered accidentally. 

However, we do not know for certain that the observed emission from all perytons follows the relation described by \Eq{eq:ch}. First of all, the very procedure of automatically searching for perytons in archival data introduced a selection bias; e.g., signals corresponding to vanishingly small $C$ and others that were not similar enough to the LB were simply ignored. (One may find this ironic, considering that the similarity between perytons and the LB was later hypothesized to be accidental.) Therefore, \Eq{eq:ch} may, in fact, reflect properties of the selection algorithm rather than an objective pattern determined by a specific physical effect. Second, even among those perytons that \textit{were} identified as such, there are some that do not quite satisfy \Eq{eq:ch}. That includes, for example, Peryton~06 in \citep{ref:burke11} and also some of the more recent observations of peryton-like signals at Bleien Observatory, Switzerland \citep{arX:sainthilaire14}. This is particularly notable considering that surveys of GHz bursts cover only a narrow frequency band ($\Delta f/f \lesssim 0.25$), thus leaving a lot of freedom for fitting. Hence we may not actually have enough evidence to conclude whether the frequency of perytons, whatever those are, follows a power scaling like \Eq{eq:ch} or, for that matter, any other universal scaling.\footnote{This is also consistent with the study \citep{arX:katz14} that indicates: if perytons were produced by terrestrial basic plasma effects leading to \Eq{eq:ch}, then the values of $C$ would have been very different from those seen in practice.} In this sense, the model of a BL as a nonstationary electromagnetic cavity seems to be generally consistent with the peryton frequency chirps that are observed.

\subsection{Observation patterns} 

One of the perytons' puzzling features has been that, although supposedly being atmospheric events, perytons do not correlate significantly with thunderstorms \citep{ref:bagchi12, ref:burke2011b}. At the first sight, this seems to distinguish perytons from BLs, for which such correlations are commonly known \citep{ref:smirnov93}. However, the distinction may be caused by another selection bias introduced by observations. BLs are typically detected visually at large viewing angles. Perytons, in contrast, are detected at small viewing angles by RF antennas, which are not even pointed at a storm by default. If perytons are indeed produced by BLs, which are rare events themselves, their generation by distant thunderstorms should be extremely hard to detect this way. As regarding strong storms that occur on-site and could produce detectable BLs, those require stowing radio telescopes\footnote{Parkes Radio Telescope Users Guide, \url{http://www.parkes.atnf.csiro.au/observing/documentation/user_guide/pks_ug.pdf}} and thus, apparently, have not been monitored. 

Hence we conclude that radiation from BLs observed accidentally by a radio telescope should not be expected to demonstrate significant correlations with thunderstorm proximity. Moreover, lightning strikes may not be the primary cause of BLs in the vicinity of a telescope in the first place. BLs may rather originate, spontaneously, from man-made electromagnetic energy on-site, where the presence of large conducting surfaces and, supposedly, powerful electric transformers makes such events more probable than in natural settings \citep{ref:smirnov93}. (The characteristic distance would be $D/\theta \sim 11\,\text{m}$ for $D \sim 20\,\text{cm}$ and $\theta \sim 1^\circ$, assuming the Parkes antenna.) If the location of such surfaces relative to the antenna is fixed, that would also explain why the signals appear almost invariably in all the telescope beams at the same time. Furthermore, since BLs typically move only at moderate speeds, $\sim 1\,\text{m}/\text{s}$ \citep{ref:smirnov93}, their emission detected by the antenna should be relatively stable on sub-second time scales, and that is precisely what is observed for perytons too.

Finally, note that, while BLs may not be related to thunderstorms directly, weather may still affect the environment properties; thus, some correlations could be anticipated between weather conditions and the occurrence of even those BLs that do not result from lightning strikes. Although the underlying physics is not understood yet, BLs are typically observed around midday and during foul weather \citep{ref:smirnov93}. That is exactly how perytons appear too, to the extent that their known statistics \citep{ref:bagchi12} can be considered representative. [Of course, the arrival times of some perytons are clustered and may not be entirely random \citep{ref:burke11, ref:kocz12}, but so can be BLs, especially if they are produced on-site.] Also note that, even under most favorable conditions, BLs still remain rare events, which is another feature that they have in common with perytons.

\section{RF emission mechanism}
\label{sec:rfmech}

Let us now discuss whether our model of perytons can explain how the RF energy is produced or confined long enough within the BL cavity. The existing RF models of BL \citep{ref:kapitsa55, ref:watson60, ref:tonks60, ref:silberg61, ref:dawson69, ref:jennison73, ref:endean76, ref:muldrew90, ref:zheng90, ref:wesselberg03} are too sketchy to answer this question, so it may be premature to speculate on specifics. On the other hand, there is a growing experimental and theoretical evidence that most of the BL energy may be accumulated in a non-RF form, namely, in the form of internal molecular excitations or chemical energy \citep{ref:paiva07, ref:dikhtyar06, ref:alexeff04, ref:bychkov02, ref:abrahamson00, ref:brandenburg98, ref:zhiltsov95, ref:golka94, ref:ohtsuki91}. Thus, a hybrid mechanism may be in effect, such that the RF power does not produce a BL but is generated as a byproduct through a ``plasma maser'' mechanism akin to that in \citep{ref:handel94}. Specifically, this could work as follows.

With the expected temperature of several thousand Kelvin \citep{ref:cen14}, the body of a BL acts as a cold plasma for RF oscillations. This means that its dielectric susceptibility exhibits temporal (but not spatial) dispersion determined by the nonzero electron density $n_e$. For waves with a given angular frequency $\omega = 2 \pi f$, the corresponding dielectric susceptibility is $\chi \approx - \omega_p^2/[\omega(\omega + i \nu)]$, where $\omega_p = (4\pi n_e e^2/m_e)^{1/2}$ is the plasma frequency, $e$ and $m_e$ are the electron charge and mass, and $\nu$ is the electron scattering rate. The scattering is mostly due to collisions with neutrals, so $\nu$ can be taken roughly as a constant, say, $\nu \sim 10^{12}\,\text{s}^{-1}$ \citep{ref:dawson69}. 

The effect of collisions is different for different waves \citep{book:alexandrov}. Electrostatic oscillations would decay at the rate $\sim \nu$ and thus are impossible in such plasma in the GHz range. But electromagnetic oscillations, whose decay rate is $\sim \nu\omega_p^2/\omega^2$, may be possible. Since $\nu \gg \omega$ in the frequency range of our interest, we can approximate
\begin{equation}
|\chi| \sim 5n_{13} f^{-1}_{\text{GHz}} \sim n_{13}.
\end{equation}
Here $f_{\text{GHz}}$ is the frequency in GHz, and $n_{13}$ is the electron density measured in units $10^{13}\,\text{cm}^{-3}$. It is feasible that the initial electron density is in the range $n_{13} \sim 1$ and is larger at the periphery, as would occur, e.g., in the case of a blast wave. Then an RF cavity is formed for electromagnetic oscillations, to which excited molecules can emit a fraction of their energy much like in the well-known hydrogen maser \citep{book:major}. (We suppose that the emission is not particularly sensitive to the cavity size, assuming that many quantum transitions can contribute; e.g., rotational energy of polymer molecules \citep{ref:bychkov02} can be involved, which naturally have a broad distribution of resonant frequencies.) Initially, the RF energy is only poorly confined in such a cavity and will dissipate rapidly, but there is a feedback mechanism that can improve the confinement, namely, as follows. 

Note that a BL is expected to consist of \textit{dusty} plasma \citep{ref:meir13}, so $n_e$ can vary significantly through absorption and release of electrons from the the dust particle surfaces. RF power is one of the determining factors here. As shown experimentally in \citep{ref:berndt06}, application of RF field can decrease $n_e$ in dusty plasma by many times. The specific nature of this effect, which is being debated \citep{ref:schweigert12}, is not important for our discussion. What is important, however, is that the effect is local and much stronger than that caused by ponderomotive expulsion \citep{ref:zheng90}. Already weak RF oscillations may then be able to substantially steepen the $n_e$ profile within the plasmoid. Hence a well-defined electromagnetic mode can form and serve as a narrow-band transmitter of RF radiation at frequency $f_c$ [\Eq{eq:fce}]. On the other hand, as the RF energy confinement improves, $n_e$ continues to decrease in the BL core, leading to the increase of $D$ and decrease of $f_c$; hence the transmission will be chirped until the maser is exhausted. 

Note that the sketch we presented here is intended only to show how one mechanism \textit{might} possibly be common to perytons and BLs. What we actually draw attention to at this point is merely that the BL expansion is seen at least in some measurements \citep{ref:cen14}, so in one way or another, chirping of BL radiation is anticipated. Also, even though most BLs are seen with approximately constant radius, these observations pertain only to long-living structures that are easier to be seen, and, should those emit in the RF range, the radiation would not be identified as perytons (Sec.~\ref{sec:chirp}). In contrast, BLs that deteriorate on the sub-second time scale characteristic of perytons are less likely to be even be noticed and, when they are noticed, are reported as transient.

\section{Discussion}

Our conjecture that two types of curious observations, perytons and BLs, actually result from one and the same phenomenon leads to two predictions. One, we predict that atmospheric BLs emit chirped GHz radiation. Two, if perytons are indeed signatures of BL, then they should also emit optical radiation. Facilities that observe perytons do not monitor these optical emissions, but maybe they should. Also note that, should the prediction of either of these emissions be confirmed, it would not only support strongly our theory that perytons and BL are coincidentally the same phenomenon, but it would also lead to the following consequences.

First, if perytons are indeed signatures of BLs, then they should have a common physical mechanism. Our proposal of such a mechanism here is only a preliminary sketch and describes one of many possibilities. However, what definitely would follow from the coincidence of perytons and BLs is that mechanisms not common to both types of observations could then be ruled out. Second, a confirmation of the coincidence of perytons and BL would suggest that other unidentified curious RF signals be reconsidered in light of this coincidence. For instance, the LB may not be an extraterrestrial signal after all, as has been already suggested \citep{arX:kulkarni14}. This also applies to the similar ``fast radio bursts'' (FRBs) reported more recently \citep{foot:frb}. The FRBs, including the LB, may be peryton-like signatures of BLs. We might also understand the so-called ``Wow!'' signal, a famous yet still-enigmatic $1.42\,\text{GHz}$ burst that was received in 1977 by the Big Ear radio telescope, Ohio, and lasted for 72\,s \citep{book:gray}. It is not unfeasible that, although not chirped, this signal is explainable as RF emission from a BL too, as large enough BLs are known indeed to last over a minute \citep{ref:smirnov93}. Thus, what we suggest here is a connection not only between BLs and perytons, but also, possibly, between these curious observations and other known GHz signals that remain unidentified.

In summary, the hypothesis is advanced here that two types of curious observations, perytons and BLs, actually result from one and the same phenomenon. Although this connection remains speculative, the circumstantial evidence is significant and leads to testable predictions, as summarized in Table~\ref{tab:tab}. We also point out that, irrespective of whether BLs and perytons are connected, how perytons are searched for in observational data may warrant reconsideration, for existing procedures may be discarding events that has the same nature as known perytons.

\acknowledgments 

The work was supported by the NNSA SSAA Grant No. DE274-FG52-08NA28553, by the U.S. DOE Contract No. DE-AC02-09CH11466, and by the U.S. DTRA Grant No. HDTRA1-11-1-0037. We are also thankful to the reviewer for useful comments.

\clearpage

\begin{table*}
\begin{center}{\scriptsize
\begin{tabular}{c @{\qquad}c @{\qquad} c}
& BLs & perytons\bigskip\\
\hline
&&\\
explanation & unclear & unclear \bigskip\\
observed patterns & midday; & midday; \\
& usually at thunderstorms, but not only & usually on rainy days, but not only \\
& & (detectors off during local storms) \\[8pt]
frequency range & predicted in the GHz range & observations limited to $\sim 1.4$\,GHz\bigskip\\
negative frequency chirp & consistent with cavity expansion & observed\\[8pt]
chirp rate $\sim -1\,\text{GHz}/\text{s}$  & predicted & observed\\
& (based on a single observation \citep{ref:cen14}) &\\[8pt]
chirping consistent with \Eq{eq:ch} & possible & assumed, but not really demonstrated \bigskip\\
duration & from a fraction of a second to a minute & fraction of a second\bigskip\\
origin & terrestrial & assumed terrestrial\bigskip \\
optical emission & observed & predicted \bigskip\\
other curious observations & consistent with larger BL & similar frequency \\
(``Wow!'' signal) & in terms of duration &\bigskip\\
\hline
\end{tabular}
}\end{center}
\caption{Summary of the parallels between BLs and perytons.}
\label{tab:tab}
\end{table*}


\begin{thebibliography}{}

\bibitem[{Alexandrov {et~al.}(1984)}]{book:alexandrov}
Aleksandrov, A. F., Bogdankevich, L. S., \& Rukhadze, A. A. 1984, 
Principles of plasma electrodynamics (New York: Springer-Verlag)

\bibitem[{Stix(1992)}]{book:stix}
Stix, T.~H. 1992, Waves in plasmas (New York: AIP) 

\bibitem[{Abrahamson \& Dinniss(2000)}]{ref:abrahamson00}
Abrahamson, J. \& Dinniss, J. 2000, Nature, 403, 519

\bibitem[{Alexeff {et~al.}(2004)}]{ref:alexeff04}
Alexeff, I., Thiyagarajan, M., \& Grace, M. 2004, IEEE Trans. Plasma Sci., 32,
  1378

\bibitem[{Bagchi {et~al.}(2012)}]{ref:bagchi12}
Bagchi, M., Nieves, A.~C., \& McLaughlin, M. 2012, Mon. Not. R. Astron. Soc.,
  425, 2501

\bibitem[{Berndt {et~al.}(2006)}]{ref:berndt06}
Berndt, J., Kova\v{c}evi\'{c}, E., Selenin, V., Stefanovi\'{c}, I., \& Winter,
  J. 2006, Plasma Sources Sci. Technol., 15, 18

\bibitem[{Brandenburg \& Kline(1998)}]{ref:brandenburg98}
Brandenburg, J.~E. \& Kline, J.~F. 1998, IEEE Trans. Plasma Sci., 26, 145

\bibitem[{Burke-Spolaor(2011)}]{phd:burke}
Burke-Spolaor, S. 2011, PhD thesis, Swinburne Univ.

\bibitem[{Burke-Spolaor {et~al.}(2011)}]{ref:burke11}
Burke-Spolaor, S., Bailes, M., Ekers, R., Macquart, J.-P., \& Crawford, F. III, 2011,
  Astrophys. J., 727, 18
  
\bibitem[{Burke-Spolaor {et~al.}(2011b)}]{ref:burke2011b}
Burke-Spolaor, S., Ekers, R., Macquart, J.-P., 2011, in \textit{General Assembly and Scientific Symposium, 2011 XXXth URSI}, DOI:10.1109/URSIGASS.2011.6051033.

\bibitem[{Bychkov(2002)}]{ref:bychkov02}
Bychkov, V.~L. 2002, Phil. Trans. R. Soc. A, 360, 37

\bibitem[{Cen {et~al.}(2014)}]{ref:cen14}
Cen, J., Yuan, P., \& Xue, S. 2014, Phys. Rev. Lett., 112, 035001

\bibitem[{Cohen {et~al.}(2005)}]{foot:esf}
  Cohen, J., Spoelstra, T., Ambrosini, R., \& van Driel, W.
  (editors), \textit{CRAF Handbook for Radio Astronomy}, 3rd edition (European
  Science Foundation, 2005), Sec.~7.1
\bibitem[{Dawson \& Jones(1969)}]{ref:dawson69}
Dawson, G.~A. \& Jones, R.~C. 1969, Pure Appl. Geophys., 75, 247

\bibitem[{Dikhtyar \& Jerby(2006)}]{ref:dikhtyar06}
Dikhtyar, V. \& Jerby, E. 2006, Phys. Rev. Lett., 96, 045002

\bibitem[{Endean(1976)}]{ref:endean76}
Endean, V.~G. 1976, Nature, 263, 753

\bibitem[{Gray(2012)}]{book:gray}
Gray, R.~H. 2012, The elusive Wow: searching for extraterrestrial intelligence
  (Chicago: Palmer Square Press)

\bibitem[{Handel \& Leitner(1994)}]{ref:handel94}
Handel, P.~H. \& Leitner, J.-F. 1994, J. Geophys. Res., 99, 10689

\bibitem[{Jennison(1973)}]{ref:jennison73}
Jennison, R.~C. 1973, Nature, 245, 95

\bibitem[{Golka(1994)}]{ref:golka94}
Golka, R. K.~G. Jr. 1994, J. Geophys. Res., 99, 10679

\bibitem[{Kapitsa(1955)}]{ref:kapitsa55}
Kapitsa, P.~L. 1955, Dokl. Akad. Nauk SSSR, 101, 245

\bibitem[{Katz(2014)}]{arX:katz14}
Katz, J.~I. 2014, Ap. J., 788, 34

\bibitem[{Khan(2014)}]{arX:khan14}
Khan, M.~D. 2014, arXiv:1404.5080.

\bibitem[{Kocz {et~al.}(2012)}]{ref:kocz12}
Kocz, J., Bailes, M., Barnes, D., Burke-Spolaor, S., \& Levin, L. 2012, Mon.
  Not. R. Astron. Soc., 420, 271

\bibitem[{Kosarev {et~al.}(1968)}]{ref:kosarev68}
Kosarev, E.~L., Vaganov, A.~B., Zakirov, B.~S., Luganskii, L.~B., Narusbek,
  E.~A., \& Samosyuk, V.~N. 1968, Zh. Tekh. Fiz., 38, 1831
  [1969, Sov. Phys. Tech. Phys., 13, 1477]\notetoeditor{Given in brackets is a reference to the English translation of the original publication.}

\bibitem[{Kosarev {et~al.}(1970)}]{ref:kosarev70}
Kosarev, E.~L., Zatsepin, V.~G., \& Mitrofanov, A.~V. 1970, J. Geophys. Res.,
  75, 7524

\bibitem[{Kulkarni {et~al.}(2014)}]{arX:kulkarni14}
Kulkarni, S.~R., Ofek, E.~O., Neill, J.~D., Zheng, Z., \& Juric, M. 2014, arXiv:1402.4766

\bibitem[{Lorimer {et~al.}(2007)}]{ref:lorimer07}
Lorimer, D.~R., Bailes, M., McLaughlin, M.~A., Narkevic, D.~J., \& Crawford, F.
  2007, Science, 318, 777

\bibitem[{Major(2007)}]{book:major}
Major, F.~G. 2007, The quantum beat: principles and applications of atomic
  clocks (New York: Springer), 2nd edition, Chap.~11.

\bibitem[{Meir {et~al.}(2013)}]{ref:meir13}
Meir, Y., Jerby, E., Barkay, Z., Ashkenazi, D., Mitchell, J.~B., Narayanan, T.,
  Eliaz, N., LeGarrec, J.-L., Sztucki, M., \& Meshcheryakov, O. 2013,
  Materials, 6, 4011

\bibitem[{Muldrew(1990)}]{ref:muldrew90}
Muldrew, D.~B. 1990, Geophys. Res. Lett., 17, 2277

\bibitem[{Ohtsuki \& Ofuruton(1991)}]{ref:ohtsuki91}
Ohtsuki, Y.~H. \& Ofuruton, H. 1991, Nature, 350, 139

\bibitem[{Paiva {et~al.}(2007)}]{ref:paiva07}
Paiva, G.~S., Pavao, A.~C., de Vasconcelos, E.~A., Mendes, O. Jr., \& da Silva, E. F. Jr. 2007, Phys. Rev. Lett., 98, 048501

\bibitem[{Saint-Hilaire {et~al.}(2014)}]{arX:sainthilaire14}
Saint-Hilaire, P., Benz, A.~O., \& Monstein, C. 2014, arXiv:1402.0664

\bibitem[{Schweigert \& Alexandrov(2012)}]{ref:schweigert12}
Schweigert, I.~V. \& Alexandrov, A.~L. 2012, J. Phys. D: Appl. Phys., 45,
  325201

\bibitem[{Silberg(1961)}]{ref:silberg61}
Silberg, P.~A. 1961, J. Appl. Phys., 32, 30

\bibitem[{Smirnov(1993)}]{ref:smirnov93}
Smirnov, B.~M. 1993, Phys. Rep., 224, 151

\bibitem[{Thornton {et~al.}(2013)}]{foot:frb}
Thornton, D., Stappers, B., Bailes, M., Barsdell, B., Bates, S., Bhat, N.
  D.~R., Burgay, M., Burke-Spolaor, S., Champion, D.~J., Coster, P., D'Amico,
  N., Jameson, A., Johnston, S., Keith, M., Kramer, M., Levin, L., Milia, S.,
  Ng, C., Possenti, A., \& van Straten, W. 2013, Science, 341, 53; also see references cited therein.

\bibitem[{Tonks(1960)}]{ref:tonks60}
Tonks, L. 1960, Nature, 187, 1013

\bibitem[{Watson(1960)}]{ref:watson60}
Watson, W. K.~R. 1960, Nature, 185, 449

\bibitem[{Wessel-Berg(2003)}]{ref:wesselberg03}
Wessel-Berg, T. 2003, Physica D, 182, 223

\bibitem[{Zheng(1990)}]{ref:zheng90}
Zheng, X.-H. 1990, Phys. Lett. A, 148, 463

\bibitem[{Zhil'tsov {et~al.}(1995)}]{ref:zhiltsov95}
Zhil'tsov, V.~A., Manykin, E.~A., Petrenko, E.~A., Skovoroda, A.~A., Leither,
  J.~E., \& Handel, P.~H. 1995, J. Exp. Theor. Phys., 81, 1072

\end{thebibliography}
\end{document}